# Quantum approach to the dynamical systems modeling

Yu.I. Bogdanov*[abc], N.A. Bogdanova[ab], D.V. Fastovets**[ab], V.F. Lukichev[a]

[a]Valiev Institute of Physics and Techonology of Russian Academy of Sciences, Russia, Moscow; [b]National Research University of Electronic Technology (MIET), Russia, Moscow; [c]National Research Nuclear University (MEPhI), Russia, Moscow

**ABSTRACT**

We present a general approach to the classical dynamical systems simulation. This approach is based on classical systems extension to quantum states. The proposed theory can be applied to analysis of multiple (including non-Hamiltonian) dissipative dynamical systems. As examples, we consider the logistic model, the Van der Pol oscillator, dynamical systems of Lorenz, Rössler (including Rössler hyperchaos) and Rabinovich-Fabrikant. Developed methods and algorithms integrated in quantum simulators will allow us to solve a wide range of problems with scientific and practical significance.

## 1. INTRODUCTION

The idea of quantum computers and quantum simulators was invented in 1982 by R. P. Feynman [1]. The modeling of quantum states is an exponential complexity class problem on classical computer. He proposed to obtain the solution of such problems using new quantum elements base: using some quantum systems to modeling others. For example, using quantum electromagnetic field state in multichannels quantum optical interference devices (optical quantum simulator) to obtain the structure of large molecular systems and new materials properties. It was shown in 1994 by Zeilinger group [2] that an arbitrary finite unitary transform can be obtained using default quantum optics devices. The universal quantum simulator circuit [3] was proposed in 1996. The first demonstration experiments with molecular energy spectrum evaluating were performed in 2010 [4].

This work is a continuation of our researches performed in [5,6]. Our developed approach is based on the extension of an arbitrary classical dynamical system to a quantum system. Note that the dynamical systems research is a fundamental stage for the management of physical, technical and socio-economic systems.

It is important to note that a quantum dynamical system contains more information than the original classical system. In particular, according to quantum mechanics, we can consider not only the coordinate representation, but also the momentum representation. We know that the mutually complementary coordinate and momentum representations are linked through the fundamental Heisenberg uncertainty relation. Compression of phase space in a certain direction in the coordinate representation indicates the extension in momentum space in the same direction, and vice versa. We use this property to find the Lyapunov exponents in the original coordinate space. We perform more simple calculations in momentum space. From the fundamental point of view, simultaneous consideration of mutually complementary coordinate and momentum frameworks provides a deeper understanding of the nature of chaotic behavior in dynamical systems. From the computational point of view, the new formalism provides a basis for the analysis of complex dynamical systems using quantum simulators.

*bogdanov_yurii@inbox.ru; **fast93@mail.ru

## 2. REPRESENTATION OF CLASSICAL DYNAMICAL SYSTEMS BY MEANS OF QUANTUM STATISTICAL ENSEMBLES

Classical dynamical system is described by means of an stationary (autonomous) system of differential equations in the following form (for example [7]):

$$\frac{dx_j}{dt} = F(x_1,...,x_n), \quad j=1,2,...,n ,  \qquad (1)$$

where $n$ is a dimension of dynamical system. Values $F(x_1,...,x_n)$ specify components of the particle's velocity as a function of the coordinates. Stationary (autonomy) means that time $t$ is not included in the $F_j$ functions which define the system's (1) right side.

Dynamical system's description in terms of statistical ensembles assumes the definition of the corresponding density distribution $\rho(x_1,...,x_n,t)$ in the phase space. Value $\rho dx_1...dx_n$ defines the probability of particle's localization in an elementary volume $dx_1...dx_n$. Usually, the probability is normalized to one:

$$\int \rho(x_1,...,x_n,t) dx_1...dx_n = 1 . \qquad (2)$$

Description in terms of statistical ensembles means that we consider the continuity equation for the density distribution instead of equations system (1). This equation can be written in traditional form:

$$\frac{\partial \rho}{\partial t} + div\mathbf{J} = 0 . \qquad (3)$$

Obviously, in our case, the vector's components are $J_j = F_j \rho$. Therefore, the continuity equation in the phase space can be written in the following form

$$\frac{\partial \rho}{\partial t} + \frac{\partial}{\partial x_j}(F_j \rho) = 0 . \qquad (4)$$

Hereafter, we imply the summation over recurring indices numbering (over index $j$ in the equation (4)). Equation (4) can be easily transformed to the analogue of the Schrodinger equation:

$$i\frac{\partial \rho}{\partial t} = L\rho, \quad L = -iF_j \frac{\partial}{\partial x_j} - i\frac{\partial F_j}{\partial x_j} , \qquad (5)$$

where $L$ is a Liouville equation. Analogues of (5) are widely used in studies on the statistical mechanics. We should note the contribution of I. Prigogine to this subject [8]. It is important to note that the Liouville operator is not Hermitian ($L \neq L^{\dagger}$), if the divergence of the velocity is different from zero ($div \vec{F} \neq 0$). On the other side, the Liouville operator is Hermitian ($L = L^{\dagger}$), if the divergence is equal to zero ($\frac{\partial F_j}{\partial x_j} = 0$ - case of incompressible flow in phase space). The Hamiltonian dynamics of the classical dynamical system gives an example of such incompressible flow. In this case, a mechanical system with $n$ degrees of freedom is represented by a statistical ensemble in a phase space with $2n$ dimension. The equations of the dynamical system are the Hamilton equations:

$$\frac{dx_j}{dt} = F_j^x = \frac{\partial H(x_1,...,x_n,p_1,...,p_n)}{\partial p_j} \quad \frac{dp_j}{dt} = F_j^p = -\frac{\partial H(x_1,...,x_n,p_1,...,p_n)}{\partial x_j} \tag{6}$$

where $j = 1, 2, ..., n$. Here, $H(x_1,...,x_n,p_1,...,p_n)$ is a Hamilton function, $F_j^x$ is a phase space velocity's component corresponding to $j$-th degree of freedom, and $F_j^p$ is a corresponding impulse component of the velocity in the phase space. Here, the velocity divergence in the phase space turns to zero, since

$$\frac{\partial F_j^x}{\partial x_j} + \frac{\partial F_j^p}{\partial p_j} = \frac{\partial}{\partial x_j}\frac{\partial H}{\partial p_j} - \frac{\partial}{\partial p_j}\frac{\partial H}{\partial x_j} = 0.$$

Therefore, the Liouville operator turns to Hermitian operator (in classical Hamiltonian systems) and is defined by the equation:

$$L = -i\left(\frac{\partial H}{\partial p_j}\frac{\partial}{\partial x_j} - \frac{\partial H}{\partial x_j}\frac{\partial}{\partial p_j}\right) \tag{7}$$

It can be shown that the continuity equation in the phase space (3) and (4) can be considered as a consequence of the Schrodinger equation for the psi-function (complex function in general case).

$$i\frac{d\psi}{dt} = H\psi, \quad H = -iF_j\frac{\partial}{\partial x_j} - \frac{i}{2}\frac{\partial F_j}{\partial x_j}. \tag{8}$$

To obtain equations (3) or (4) from the Schrodinger equation, we need to multiply the equation for the $\psi$ function by the complex conjugate function $\psi^*$, multiply the equation for the $\psi^*$ function by the function $\psi$ and then subtract two obtained equations. It is necessary to note that $\rho = |\psi|^2$. It is important to note that the Hamiltonian in the Schrodinger equation is Hermitian.

Equation (8) that takes form of the Schrodinger equation can be considered as the basis for a natural extension of the formalism of dynamical systems to the quantum domain. All properties and characteristics of the dynamical system can be obtained from the solution of the Schrodinger equation (8). At the same time, the methods and approaches developed in quantum mechanics can be applied to the research of the Schrodinger equation (8). Note also that the dynamics of the quantum system (8) is richer than the dynamics of its classical analogue (5). In particular, we can consider some other representations except coordinate representation (the impulse representation for example). From a computational view point, the solution of the equation (8) can be obtained on a completely new element base (quantum computers and quantum algorithms aimed to solving the Schrodinger equation [9]).

If we introduce the momentum operators in the coordinate representation $P_j = -i\frac{\partial}{\partial x_j}$ then the Hamiltonian (8) can be represented in the following form:

$$H = -iF_j\frac{\partial}{\partial x_j} - \frac{i}{2}\frac{\partial F_j}{\partial x_j} = \frac{1}{2}\left(F_j P_j + P_j F_j\right). \tag{9}$$

Note that the Hamiltonian is linear in momentum. This is a significant difference from the Hamiltonians considered in traditional non-relativistic quantum mechanics.

## 3. RESEARCH OF THE LOGISTIC EQUATION USING QUANTUM THEORY METHODS

The logistic model is a simple example of complementing an arbitrary classical dynamical system to a quantum system. This system is reduced to a single-mode quantized electromagnetic field in a nonlinear medium. In this case, a stable evolution of the quantum ensemble corresponds to the electromagnetic field compression along the coordinate (the momentum expansion is occurs). The quantum analogue of the system can be investigated in an arbitrary equivalent representations (in particular, the coordinate and momentum representations).

The logistic equation (also known as the Verhulst equation) has the following form:

$$\frac{dx}{dx} = \alpha x - \beta x^2 + q. \tag{10}$$

Historically, this equation obtained as a model for describing the population size and a different populations sizes in ecological and biological systems.

We assume that, $\alpha > 0$, $\beta > 0$ and $q$ has an arbitrary sign. The first term describes the reproduction rate of the population (it is believed that the size's increase per unit time is proportional to the population size). The exponential growth of the population size is limited by the second non-linear term. For example, for $q = 0$ we can represent the right side of the equation in the form $\alpha x - \beta x^2 = (\alpha - \beta x)x$. The influence of the non-linear term can be interpreted as limited resources leading to the slow down in population growth (instead of $\alpha$ we have $(\alpha - \beta x)$; the growth rate becomes zero for $x = \alpha/\beta$ and is negative for $x > \alpha/\beta$). Finally, the third term describes a constant outside influx into the population (with $q > 0$ "population" resides from other areas in the ecosystem; with $q < 0$ "population" leaves this ecosystem). The right side of the equation can be considered as speed $F(x)$:

$$\frac{dx}{dt} = F(x), \quad F(x) = \alpha x - \beta x^2 + q. \tag{11}$$

In terms of quantum theory, such dynamical system can be described using Schrodinger equation (8) for the psi-function with the Hamiltonian (9). Note that, momentum operator $p$ doesn't commute with an operator $F(x)$ in (9). We have investigated the logistic model in three different equivalent ways: in terms of quantum optics formalism (using the creation and annihilation operators); in terms of coordinate representation; in terms of momentum representation (in the last two cases, partial differential equations were solved).

The coordinates and momentum operators can be expressed in terms of the annihilation and creation operators using the usual equations of quantum optics:

$$x = \frac{\left(a + a^\dagger\right)}{\sqrt{2}}, \quad p = -i\frac{\left(a - a^\dagger\right)}{\sqrt{2}}. \tag{12}$$

The $H$ operator, expressed using the annihilation $a$ and creation $a^\dagger$ operators, provides the evolution of the system in the Fock representation. The corresponding solution of the Schrodinger equation can be expressed in matrix exponent form:

$$\psi(t) = \exp(-iHt)\psi_0, \tag{13}$$

where $\psi_0$ and $\psi(t)$ - system's state-vector at the initial moment of time and at time $t$ respectively.

Fock representation can be transformed to other representations (in particular, to coordinate and momentum representations). The momentum operator in the coordinate representation is $p=-i\frac{\partial}{\partial x}$. In this case, the Schrodinger equation is (8). Population's density is given by the usual formula of quantum mechanics: $\rho(x,t)=|\psi(x,t)|^2$. The considered problem is reduced to solving the following partial first order differential equation in time and coordinate

$$\frac{\partial \psi(x,t)}{\partial t}+F(x)\frac{\partial \psi(x,t)}{\partial x}+\frac{1}{2}\frac{\partial F(x)}{\partial x}\psi(x,t)=0 \qquad (14)$$

This equation was solved in two different ways: using difference scheme and using the Hamilton-Jacobi formalism also.

The coordinate operator in the momentum representation is $x=i\frac{\partial}{\partial p}$. The relationship between the coordinate and momentum wave functions is given by the following well-known formulas defining the direct and inverse Fourier transforms:

$$\psi(x)=\frac{1}{\sqrt{2\pi}}\int \tilde{\psi}(p)\exp(ipx)dp, \quad \tilde{\psi}(p)=\frac{1}{\sqrt{2\pi}}\int \psi(x)\exp(-ipx)dx. \qquad (15)$$

Schrodinger equation (the momentum representation) has the first order in time and the second order in coordinate

$$i\frac{\partial \tilde{\psi}(p,t)}{\partial t}=\beta p\frac{\partial^2 \tilde{\psi}(p,t)}{\partial p^2}+\beta\frac{\partial \tilde{\psi}(p,t)}{\partial p}+i\alpha p\frac{\partial \tilde{\psi}(p,t)}{\partial p}+i\frac{\alpha}{2}\tilde{\psi}(p,t)+qp\tilde{\psi}(p,t). \qquad (16)$$

The density distribution dynamics is presented on Fig. 1. Models parameters: $\alpha=0.5$, $\beta=0.1$, $q=2$.

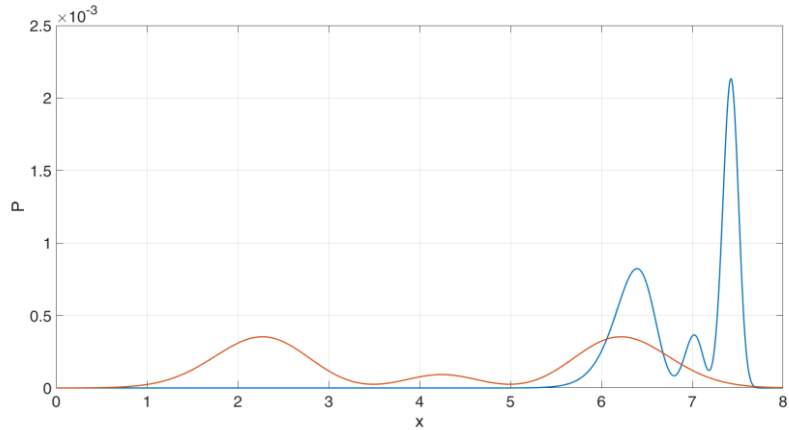

Figure 1. Density distribution dynamics in the logistic model. Blue line corresponds to $t=2$, red line corresponds to $t=0$.

At the initial time, the quantum state was modeled by superposition of harmonic oscillator modes (Chebyshev-Hermite basis):

$$\psi(x,t)=\sum_{j=0}^{4}c_j\varphi_j(x,t),$$

where $c_0 = c_2 = c_4 = 1/\sqrt{5}$, $c_1 = c_3 = i/\sqrt{5}$. Source Chebyshev-Hermite basis parameters at $t = 0$ are: $\mu = 3\sqrt{2}$, $\sigma = \dfrac{1}{\sqrt{2}}$.

## 4. VAN DER POL OSCILLATOR

Van der Pol oscillator is an example of a self-oscillating system. Such system is described by the following equation:

$$\ddot{x} + \varepsilon f(\dot{x}) + x = 0, \quad f(\dot{x}) = -\dot{x} + \frac{1}{3}\dot{x}^2. \tag{17}$$

Here, $x$ is the coordinate of the moving single mass point. This coordinate depends on time $t$. $\varepsilon$ is a coefficient characterizing non-linearity and damping force oscillations simultaneously. If we compare the equation (17) with the damped oscillations, we see that Van der Pol equation has a negative damping factor at low speed values (first term is prevail). This leads to the oscillations buildup. On the other side, at high speed values the Van der Pol equation corresponds to damped oscillations (second term prevails). The balance between these two opposite tendencies corresponds to the undamped self-oscillations (so-called limit cycle). Such effect is illustrated in Fig. 2. Models parameters: $\varepsilon = 2$, $x$ - coordinate (horizontal axis), $y = \dot{x}$ - velocity (vertical axis). The first (blue) curve describes the growth of oscillations, initially having small amplitude. The second (red) curve describes the partial oscillations damping that initially had too high amplitude. The oscillations quickly attain to the same limit cycle in the both cases.

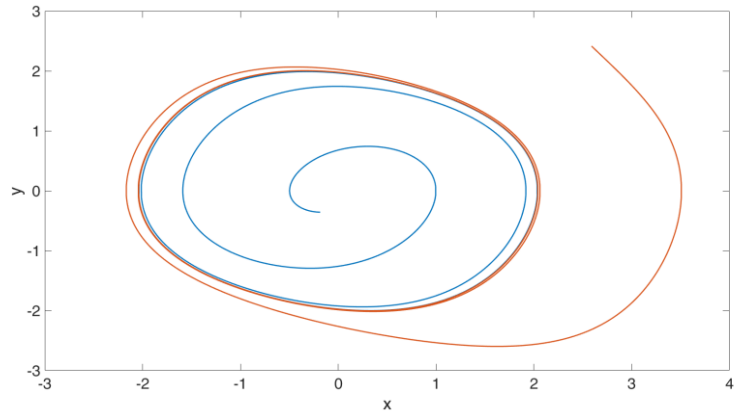

Figure 2. Phase portrait of the Van der Pol oscillator.

On second-order equation reduces to two first-order equations in term of the new variables:

$$\dot{x} = y; \tag{18}$$

$$\dot{y} = -x + \varepsilon\left(y - \frac{y^3}{3}\right). \tag{19}$$

Differentiating equation (19) by time, we can obtain

$$\ddot{y} - \varepsilon\left(1 - y^2\right)\dot{y} + y = 0. \tag{20}$$

The resulting equation (20) is often used as the initial definition of the Van der Pol oscillator instead of equation (17). The equations systems (18)-(19) can be written in the single matrix form equation for the column-vector.

$$\overline{x} = \begin{pmatrix} x \\ y \end{pmatrix}: \quad \dot{\overline{x}} = A\overline{x}, \quad A = \begin{pmatrix} 0 & 1 \\ -1 & \varepsilon\left(1 - \dfrac{y^2}{3}\right) \end{pmatrix} \quad (21)$$

Two dependencies are presented on Fig. 3. The first dependence $T(\varepsilon)$ shows the relationship between the period of self-oscillations (limit cycle) and the nonlinearity parameter $\varepsilon$. The second dependence $\lambda_n(\varepsilon)$ shows a similar relationship between the negative Lyapunov exponent and parameter $\varepsilon$. The negative Lyapunov exponent provides the compression of the phase volume across the phase trajectory (the Lyapunov exponent along the trajectory is zero).

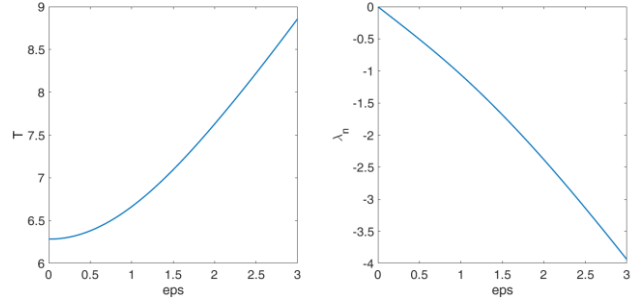

Figure 3. Dependence of the self-oscillations period (left) and the negative Lyapunov exponent (right) on the nonlinearity parameter.

Note that, $\lambda_n(\varepsilon)$ has a linear approximation $\lambda_n(\varepsilon) \approx -\varepsilon$ for small values of nonlinearity parameter.

The following figures (4a and 4b) show the first four basis functions of a two-dimensional oscillator. The first figure corresponds to $t = 0$ (the original oscillator basis, $\varepsilon = 0$). The second figure describes the basis changed under the Van der Pol unitary transformation ($\varepsilon = 1, t = 1$). Let the initial quantum state be a superposition of the first four basis functions of a two-dimensional oscillator:

$$\psi(x,y,t) = \sum_{j=0}^{3} c_j \varphi_j(x,y,t).$$

Therefore, all dynamics will be determined by the evolution of the basis functions (Fig. 4). Decomposition coefficients don't change over time.

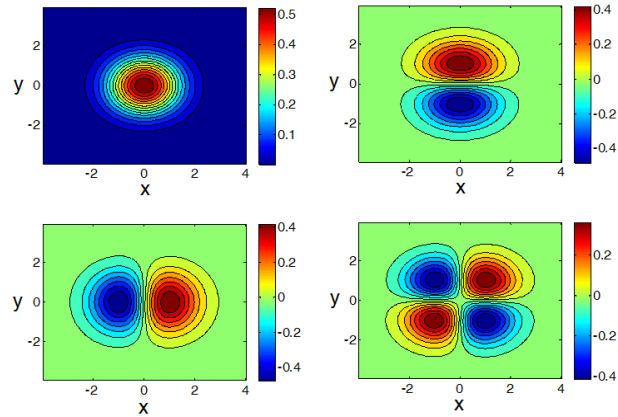

Figure 4a. The original basis of the two-dimensional oscillator (the first four functions).

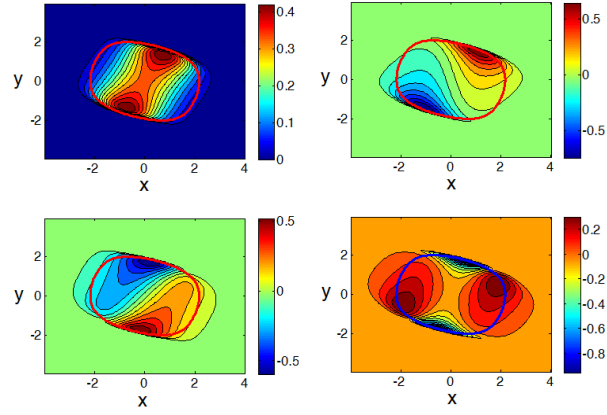

Figure 4b. The basis of the two-dimensional oscillator, changed under the unitary Van der Pol transformation (the first four functions). The solid line define the limit cycle.

Note that, the modes localize (over time) near the limit cycle curve.

## 5. EXAMPLES OF DYNAMICAL CHAOTIC SYSTEMS

Further in this article we will consider some three-dimensional and four-dimensional dynamical systems. Liouville operator (5) is not Hermitian for such systems. In this case, motion in the phase space doesn't reduce to the motion of incompressible fluid. Liouville theorem (phase volume conversation) doesn't hold also. Below, we will see that the Schrodinger evolution (8) can have a dynamical chaotic character.

We will consider four important dynamical systems: Lorenz system [10], Rössler system [11] (including Rössler hyperchaos [12]) and Rabinovich-Fabrikant system [13]. Note that, we have already considered the first two systems [5,6]. The third and fourth systems are considered for the first time in this article.

Velocity components for the Lorenz systems are:

$$F_1 = -\sigma x_1 + \sigma x_2, \quad F_2 = (r - x_3)x_1 - x_2, \quad F_3 = x_1 x_2 - b x_3. \quad (22)$$

We assume that the model parameters have the following values: $\sigma = 10$, $r = 28$, $b = 8/3$ (most often used values in numerical experiments). The Lorenz system originally appeared as the first non-trivial approximation for the problem of fluid's convection in a flat layer (cellular convection [10]). But Lorenz system can be used in other different tasks (for example, the single-mode laser model [14]).

The Rössler model is useful to modeling the equilibrium states in some chemical reactions with mixing ("three-dimensional mixer" model [15]). Rössler also introduced the four dimensional generalization of his model [12]. Hyperchaos may occur in the general model (hyperchaos is a chaotic model with two or more positive Lyapunov exponents). The hyperchaos system was introduced to describe the "three-dimensional mixer" model with a fourth substance – catalyst [12]. The velocity components of a three-dimensional system are:

$$F_1 = -x_2 - x_3, \quad F_2 = x_1 + a x_2, \quad F_3 = (x_1 - c)x_3 + b. \quad (23)$$

In this case, we will also use "default" models parameter (introduced in [11]): $a = 0.2$, $b = 0.2$, $c = 5.7$. Similarly, the equations for the four-dimensional system are:

$$F_1 = -x_2 - x_3, \quad F_2 = x_1 + a x_2 + x_4, \quad F_3 = x_1 x_3 + b, \quad F_4 = -c x_3 + d x_4. \quad (24)$$

Here, we will use the parameters values considered by Rössler in hypechaos appearance conditions [12]: $a = 0.25$, $b = 3$, $c = 0.5$ and $d = 0.05$.

The equations of the Rabinovich-Fabrikant model describe waves stochastic self-modulation in non-equilibrium media [13]. These equations can be used to describe the processes in liquid (or gaseous) medium with Tollmien-Schlichting waves [16] appearance, or the propagation of Langmuir waves in plasma [17]. The velocity components for such system are:

$$F_1 = \gamma x_1 + \left(x_3 - 1 + x_1^2\right)x_2, \quad F_2 = \left(3x_3 + 1 - x_1^2\right)x_1 + \gamma x_2, \quad F_3 = -2(\alpha + x_1 x_2)x_3. \quad (25)$$

Here, we will use the "default" parameters (obtained in [13]): $\gamma = 0.87$, $\alpha = 1.1$. The phase portraits of the coordinate and momentum representation of the Rabinovich-Fabrikant system are presented on Fig. 5.

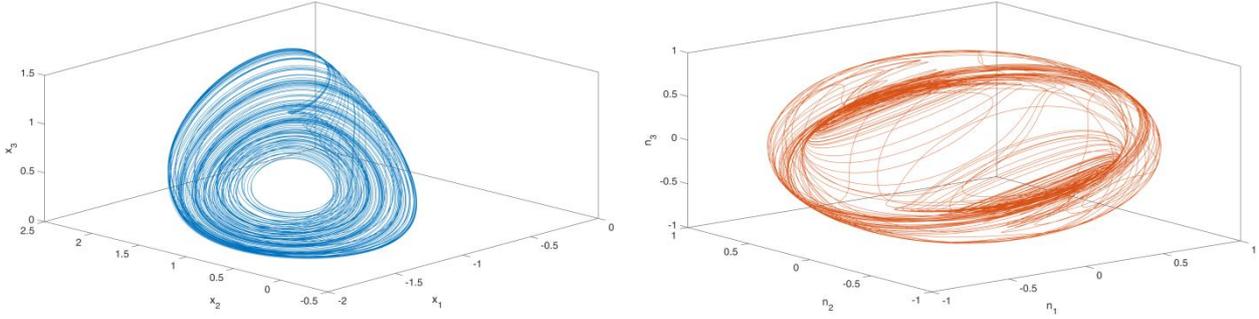

Figure 5. The Rabinovich-Fabrikant attractor. Complementary coordinate (left) and momentum (right) representations.

All presented "default" parameters of systems provide the chaotic behavior. Their attracting set is a strange attractor. In particular, for the Lorenz system $\frac{\partial F_j}{\partial x_j} = -(\sigma + 1 + b) < 0$. This implies the volume rapidly decreasing (exponentially) over time. We have a similar situation in the Rabinovich-Fabrikant model: $\frac{\partial F_j}{\partial x_j} = -2(\alpha - \gamma) < 0$. Divergence has the following non-constant value for the Rössler system: $\frac{\partial F_j}{\partial x_j} = a - c + x_1 < 0$ ($\frac{\partial F_j}{\partial x_j} = a + d + x_1 < 0$ for the four-dimensional model).

## 6. LYAPUNOV EXPONENTS FOR LORENZ, RÖSSLER AND RABINOVICH-FABRIKANT SYSTEMS

The Lyapunov exponents spectrum characterizes the stability of the statistical ensemble in the phase space. Let there be a given statistical ensemble near the point $x(t_0)$ ($t_0$ - fixed time value) in $n$-dimensional phase space. It is convenient to assume that ensemble (in coordinate representation) is characterized by a $n$-dimensional spherically symmetric Gaussian distribution with some small isotropic variance $D_{x_0} = \sigma_{x_0}^2$. In our cases, the Heisenberg inequality for the dispersions product turns into equality. Therefore, the initial momentum dispersion in any direction is equal to $D_{p_0} = \sigma_{p_0}^2 = 1/(4D_{x_0})$. Evolution will lead to the distribution's distortion (the spherical Gaussian distribution will gradually become ellipsoidal). At the same time, a similar evolution is applied to the momentum distribution. If the distribution coordinate size decreases exponentially with Lyapunov exponent $-\lambda$ ($\sigma_x(t) = \sigma_{x_0} \exp(-\lambda(t-t_0))$), then the distribution momentum size will also exponentially increase in the same direction with the exponent $\lambda$ ($\sigma_p(t) = \sigma_{p_0} \exp(-\lambda(t-t_0))$). Note that, the dynamics calculation in the momentum space is much simpler than the calculation of the initial coordinate dynamics. Lyapunov exponents are determined locally by the matrix $B$ properties $B_{jk} = -\frac{\partial F_k}{\partial x_j}$ [5,6]. The total Lyapunov exponents must be calculated by local parameters averaging over the attractor. It is important to note that time averaging is equivalent to averaging over an ensemble (ergodic hypothesis [18]). Note that we developed the method of the Lyapunov exponents calculation. This method is much simpler that the default approach (described in [7], for example). Let us consider the application of the developed approach to strange attractors (in Lorenz, Rössler and Rabinovich-Fabrikant models).

The Lyapunov exponents sum must be negative due to the dissipativity of considered systems. The attractor is an attracting set of zero-meausre in the phase space. All points localize on the attractor curve with exponential speed.

The results of numerical calculations of Lyapunov exponents for the Lorenz attractor are:

$$\lambda_1 = 0.90628 \pm 0.00068, \lambda_2 = 0.0000056 \pm 0.00015, \lambda_3 = -14.57296 \pm 0.00071$$

Obtained results are close to the values obtained by S.P. Kuznetsov [7] ($\lambda_1 = 0.897$, $\lambda_3 = -14.563$) and J.C. Sprott [19] ($\lambda_1 = 0.906$, $\lambda_3 = -14.572$). The results obtained for the Rössler Lyapunov exponents are

$$\lambda_1 = 0.07114 \pm 0.00062, \lambda_2 = 0.000016 \pm 0.00013, \lambda_3 = -5.3938 \pm 0.0018$$

Numerical values of the Lyapunov exponents for the Rössler system are close to J.C. Sprott [19] results ($\lambda_1 = 0.0714$, $\lambda_3 = -5.3943$). The hyperchaos modeling results are:

$$\lambda_1 = 0.11294 \pm 0.00029, \lambda_2 = 0.01972 \pm 0.00071, \lambda_3 = 0.000082 \pm 0.00013, \lambda_4 = -24.983 \pm 0.129$$

This results are close to J.-M. Malasoma [20] results ($\lambda_1 = 0.112$, $\lambda_2 = 0.019$, $\lambda_4 = -25.188$). And the Lyapunov exponents for the Rabinovich-Fabrikant model are:

$$\lambda_1 = 0.19872 \pm 0.00018, \lambda_2 = 0.0000056 \pm 0.00015, \lambda_3 = -0.65871 \pm 0.00018$$

Our results are similar with the results obtained by J.C. Sprott [19] ($\lambda_1 = 0.1981$, $\lambda_3 = -0.6581$).

Lorenz, Rössler and Rabinovich-Fabrikant attractors are fractals [7]. Kaplan-Yorke fractal dimension (for three- and four-dimensional systems) is given by the following equations [21,22]:

$$d_{KY} = 2 + \frac{\lambda_1}{|\lambda_3|}; \qquad (26)$$

$$d_{KY} = 3 + \frac{\lambda_1 + \lambda_2}{|\lambda_4|}. \qquad (27)$$

The calculations lead to the following results for the fractal dimension:

$d_{KY} = 2.062$ - Lorenz attractor;

$d_{KY} = 2.013$ - Rössler attractor;

$d_{KY} = 3.0053$ - four-dimensional Rössler attractor;

$d_{KY} = 2.302$ - Rabinovich-Fabrikant attractor;

## 7. CONCLUSIONS

The method of an arbitrary classical dynamical system extension to the quantum system was developed using the basic examples: the logistic model, the Van der Pol oscillator, Lorentz system, Rössler system (including Rössler hyperchaos) and Rabinovich-Fabrikant system. The Schrodinger equation was explored using the Hamilton-Jacobi mathematical formalism. Along with the initial coordinate dynamic we consider the attached dynamics corresponding to the momentum space. Simultaneous consideration of mutually complementary coordinate and momentum portraits provides a deeper understanding of the dynamical systems chaotic behavior. The new formalism simplifies the procedure of the Lyapunov exponents calculation.

## ACKNOWLEDGMENTS

The work is supported by Russian Science Foundation (RSF), project no: 14-12-01338П.